\UseRawInputEncoding
\documentclass[aps,pre,superscriptaddress,twocolumn]{revtex4-2}  % for PR
\usepackage{epsfig}
\usepackage{latexsym}
\usepackage{amsmath}
\usepackage{amsfonts}
\usepackage{amssymb}
\usepackage{amsbsy}
\usepackage{subfigure}
\usepackage{bm}
\usepackage{color}
\usepackage{dcolumn}
\usepackage{hyperref}
\usepackage[normalem]{ulem}

\graphicspath{{./figures/}}

\newcommand{\lla}{\left\langle}
\newcommand{\rra}{\right\rangle}

\newcommand{\mr}[1]{\textcolor{black}{#1}}

\begin{document}
\title{Universal swelling of tangentially driven
ring polymers in two dimensions}
\author{Antonio Lamura}
\email[]{Corresponding author, antonio.lamura@cnr.it}
\affiliation{
Istituto Applicazioni Calcolo, Consiglio Nazionale delle Ricerche (CNR),
Via Amendola 122/D, 70126 Bari, Italy}
\author{Marisol Ripoll}
\email[]{m.ripoll@fz-juelich.de}
\affiliation{Theoretical Soft
Matter and Biophysics, Institute for Advanced Simulation,
Forschungszentrum J\"{u}lich, 52428 J\"{u}lich, Germany
}
\date{\today}
\begin{abstract}
The interplay of tangential activity and excluded-volume interactions in ring polymers adsorbed to a surface, consistently results in the overall swelling of the ring configurations. This is in strong contrast to the case for three dimensional linear and ring polymers, where activity induces frequently collapsed structures. 
By means of Brownian Dynamic simulations, we investigate how the scaling properties of such active rings can be universally characterized by an activity-dependent Flory exponent, as a generalization of the equilibrium behavior. 
At high activity, an effective persistence length characterizes the conformations of active flexible rings. 
\end{abstract}
\maketitle
\newpage

\section{Introduction} 

Active matter has attracted a largely raising interest over the past decades due to its ability of exhibiting  single-particle and collective behaviors fundamentally different from those observed at equilibrium~\cite{marc:13,elge:15,bech:16,vrug:25}. Active polymers represent a particularly rich class, where internal driving, conformational flexibility, and interactions give rise to a wide range of non-equilibrium phenomena~\cite{elge:09,jiang:14,ghos:14,isele:15,isele:16,wink:17,eise:17,prathyusha:18,duman:18,anand:18,fogl:19,wink:20,eise:22,vati:24,janz:24,janz:25.2}. In particular, polymers subject to polar active forces, where each segment is propelled in a well-defined direction along the local backbone, have been shown to display a large number of behaviors such as activity-induced swelling, breakdown of self-similarity, coil-to-globule transitions, and shear-induced conformational changes~\cite{thak:11,bianco:18,faze:23,teje:24,oller:25,sapp:25,wink:25,pand:25,pand:25.2,lamu:26}.

Ring polymers provide a natural framework to investigate the role of topology in such non-equilibrium settings. Their closed architecture, lacking free ends, imposes global constraints that strongly affect both conformational and dynamical properties. These features can be relevant for the design of synthetic macromolecules or structures, also for a broad range of biological systems, including circular DNA, viral genomes, chromatin loops, actomyosin rings, and even filamentous organisms such as
microbial rings, where closed or looped geometries emerge either structurally or functionally~\cite{schl:92,kawa:08,rosa:11,ito:14,pear:18,lieb:18,tu:20,wu:19,alex:20,keya:20,schr:25,uber:25}. At equilibrium, self-avoiding rings in two dimensions are known to exhibit universal scaling properties governed by excluded-volume interactions, typically described by the Flory exponent~\cite{norm:92,witz:08,drub:10,saka:10,kuma:26,elef:26}. When activity is introduced, however, the interplay between tangential driving, excluded volume, and topology leads to qualitatively new behaviors~\cite{smerk:20,loca:21,phil:22,jain:23,kuma:23,kuma:24,wink:24,lamu:24,ayou:26}.

Recent studies have begun to address the properties of active ring polymers. In three dimensions, both active Brownian and tangentially driven models have shown that activity can significantly modify conformational fluctuations and often promotes collapsed or compact structures, depending on chain length and activity strength~\cite{mous:19,loca:21,phil:22}. In contrast, the behavior of active rings in two dimensions remains much less explored, despite its experimental relevance for systems such as microtubules or filaments confined to surfaces~\cite{kawa:08,liu:11,ito:14,keya:20} . In two dimensions, it has been shown that enhanced excluded-volume interactions and restricted configurational space 
play a crucial role, leading to behaviors that differ markedly from their three-dimensional counterparts~\cite{lamu:24}.

In this work, we investigate self-avoiding ring polymers adsorbed on a surface and driven by tangential active forces, explicitly accounting for excluded-volume interactions. 
Using Brownian Dynamics simulations \mr{in the overdamped regime}, we show that the interplay between activity and self-avoidance consistently leads to a universal swelling of ring conformations over a broad range of activity strengths and polymer lengths. This finding is in strong contrast with the collapse observed in three-dimensional active polymers. This difference is known to emerge due to the reduced dimensionality, since in the absence of excluded volume interactions two-dimensional polar active polymers have also shown collapsed states~\cite{lamu:24}. We demonstrate that the conformational statistics can be described by an activity-dependent Flory exponent, providing a unified scaling framework for passive and active regimes. Furthermore, at high activity, the conformations of flexible rings can be characterized by an effective persistence length, indicating the emergence of activity-induced stiffness at large scales.
The activity-dependent Flory exponent universally applies to all investigated quantities, playing therefore a similar role as that in equilibrium.

\section{Simulation model} \label{sec:model}

A closed ring is considered in two spatial dimensions consisting of $N$ beads of mass $M$. Connected beads interact via the harmonic potential 
\begin{equation}\label{bond}
U_{bond}=\frac{\kappa_h}{2} \sum_{i=1}^{N}
\left(|{\bm r}_{i+1}-{\bm r}_{i}|-l\right)^2 ,
\end{equation}
where $\kappa_h$ is the spring constant,
${\bm r}_i$ denotes the position vector of the $i-$th bead
($i=1,\ldots,N$) with the notations ${\bm r}_{N+1}={\bm r}_1$
and ${\bm r}_{0}={\bm r}_N$, and $l$ is the average bond extension.
Excluded-volume interactions between non-bonded beads are
introduced by the truncated and shifted Lennard-Jones potential
\begin{equation}
U_{ex} =
4 \epsilon \left[ \left(\frac{\sigma}{r}\right)^{12}
-\left(\frac{\sigma}{r}\right)^{6} +\frac{1}{4}\right] \Theta\left(2^{1/6}\sigma -r\right) ,
\label{rep_pot}
\end{equation}
where $\epsilon$ is the volume-exclusion energy,
$r$ is the distance between two non-connected beads, and
$\Theta(x)$ is the Heaviside function ($\Theta(x)=0$ for $x<0$ and
$\Theta(x)=1$ for $x \ge 0$). While the harmonic potential maintains the ring topology, the excluded-volume interaction ensures that chain self-intersection is avoided.
 
The tangential active force ${\bm F}_i^a$ is applied on each bead
by assuming a constant force along the bonds direction, {\em i.e.} $f^a ({\bf r}_{i}-{\bf r}_{i-1})/l$%(i=1,\ldots,N)$ 
~\cite{jiang:14}.
This force is equally distributed between the adjacent beads so that the net force acting on the $i$-th bead is ~\cite{jiang:14,isele:15,anand:18,phil:22,lamu:24} 
\begin{equation}
{\bm F}_i^a =
\frac{f^a}{2 l} \left ( {\bm r}_{i+1}-{\bm r}_{i-1} \right ) , i=1,\dots,N.
\label{force}
\end{equation}
The force modulus $|{\bm F}_i^a|$
varies between $0$, when the two consecutive bond vectors ${\bm t}_{i}={\bm r}_{i+1}-{\bm r}_{i}$
and ${\bm t}_{i-1}={\bm r}_{i}-{\bm r}_{i-1}$ are antiparallel,
and $f^a$, when they are parallel, since the bond length is constrained to be
$l$. 
Moreover, the forces in (\ref{force}) sum up to zero along the chain \cite{phil:22}.
The strength of the active force is quantified by 
the P\'eclet number
$\mathrm{Pe} =f^a l / (k_B T)$ \cite{loca:21,teje:24}, which is independent on the
chain length,
where $k_B T$ is the thermal energy, $T$ is the temperature, and $k_B$ is
Boltzmann's constant.
Newton's equations of motion of beads are integrated by
using the velocity-Verlet algorithm with time step $\Delta t_p$
\cite{swop:82,alle:87}.
Alternative implementations of the activity, such as a constant force along each bond~\cite{bianco:18,loca:21,mira:23,janz:25}, do not show a significant difference as we checked. Models referring to motility-assay-driven polymers~\cite{roy:26b,roy:26} with a force-dependent response of the motor proteins, might result in a contraction of the ring polymer in two dimensions. 

The ring is immersed in a Brownian heat bath that is modeled by the Brownian multiparticle collision (MPC) method \cite{kiku:03,ripo:07,gomp:09,lamu:24_2}. 
In this method, every bead interacts with $\rho$ ghost solvent particles of mass $m$ in order to mimic the interaction with a fluid volume, which is performed by one collision with an effective particle having momentum sampled from a Maxwell-Boltzmann distribution with zero mean and variance $\rho m k_B T$.
Using the stochastic rotation dynamics of the MPC method~\cite{ihle:01,lamu:01,gomp:09},
collisions between polymer beads and virtual solvent particles consist of randomly rotating
the relative velocity of each polymer bead, with respect to the
center-of-mass velocity of the bead itself and the corresponding fluid
particle, at angles $\pm \alpha$. 
Collisions are executed at time intervals $\Delta t$, with   $\Delta t > \Delta t_p$. 
This method allows an eventual direct comparison with MPC where hydrodynamic interactions would be included, and in this random version it is efficient and equivalent to standard Brownian Dynamics. \mr{It has been recently shown~\cite{lamu:26} that the evolution equation of beads of mass $M$ takes the form of a discrete underdamped Langevin equation where the explicit expression of the friction coefficient $\xi$ is provided.} 

Simulations are carried out with the choices $\alpha=130^{o}$,
$\Delta t=0.1 t_u$, with time unit $t_u=\sqrt{m l^2/(k_B T)}$,
$M= \rho m$ with
$\rho=5$, $\kappa_h l^2/(k_B T)=10^4$, $\epsilon/(k_B T)=1$,
$\sigma/l=1$,
and $\Delta t_p=10^{-2} \Delta t$.
The value of $\kappa_h$ ensures that bond length fluctuations are negligible in any non-equilibrium condition.
\mr{The friction coefficient results  $M/\xi \approx 0.1t_u$, such that inertial effects are already damped out on the scale of one simulation time, and do not play any role in our simulations.}

Rings with $N=50, 100, 200, 400, 600$ are here considered with an active 
force $f^a$ varied within the interval $0 \leq \mathrm{Pe} \leq 40$ 
of P\'eclet number while still keeping constant
the value of the bond length \cite{teje:24}.
Chains are initially set in a circular shape and let
equilibrate up to time $10^6 t_u$. 
Data are accumulated in runs of duration $10^7 t_u$, much larger than $T$,
the active tank-treading period, which has shown to decrease with the intensity of the applied active force as $T \propto (N/\mathrm{Pe}) t_u$ \cite{lamu:24}.

\section{Results} \label{sec:results}

\begin{figure}[ht!]
\includegraphics[width=.48\textwidth,angle=0]{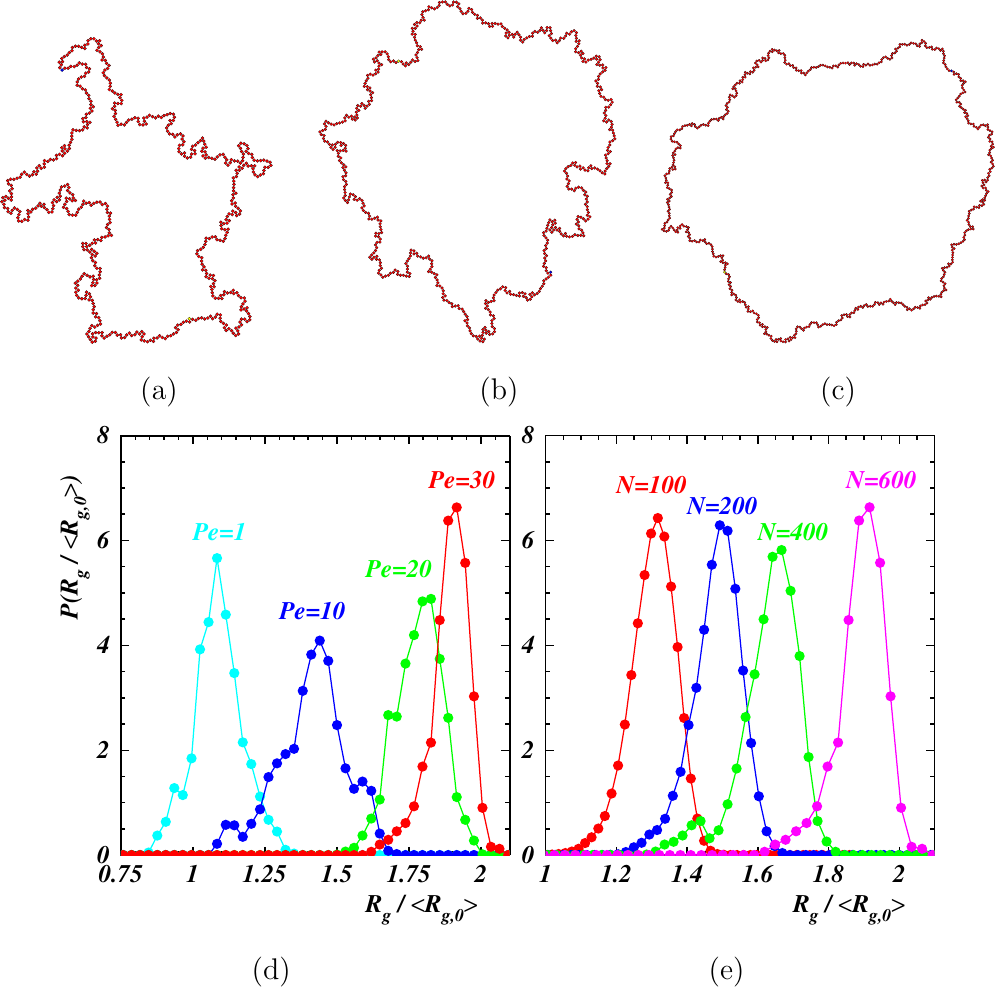}
\caption{Swelling of tangentially driven ring polymers in two dimensions. 
a-c)~Snapshots of ring polymers with $N=600$ beads and increasing activity, a)~Pe~$=1$, b)~Pe~$=10$, c)~Pe~$=30$. 
d),e)~Normalized probability distribution function of the radius of gyration, $R_g$, normalized by the value at equilibrium $\lla R_{g,0}\rra$, with d)~$N=600$ and varying Pe, e)~Pe $=30$ and varying~$N$. 
Symbols correspond to simulation results with colored lines as guides to the eye.
\label{fig:conf}
}
\end{figure}

In the two dimensional case here investigated, polymer rings show to monotonously increase their swelling with increasing directional active force. 
Characteristic ring configurations are depicted in Fig.~\ref{fig:conf} a-c for the longest simulated cyclic
polymer.
To precisely quantify the degree of swelling, we first compute the radius of gyration of the rings $R_g$, defined as $R_g=\sqrt{\sum_i ({\bm r}_{i}-{\bm r}_{cm})^2/N}$ where ${\bm r}_{cm}=\sum_i {\bm r}_{i}/N$ denotes the ring center of mass. 
This is compared with the value of the radius of gyration in equilibrium $\lla R_{g,0}\rra$, calculated for each ring size as an ensemble average, indicated by $\lla \cdots \rra$. 

\begin{figure*}[ht]
\includegraphics*[width=.99\textwidth,angle=0]{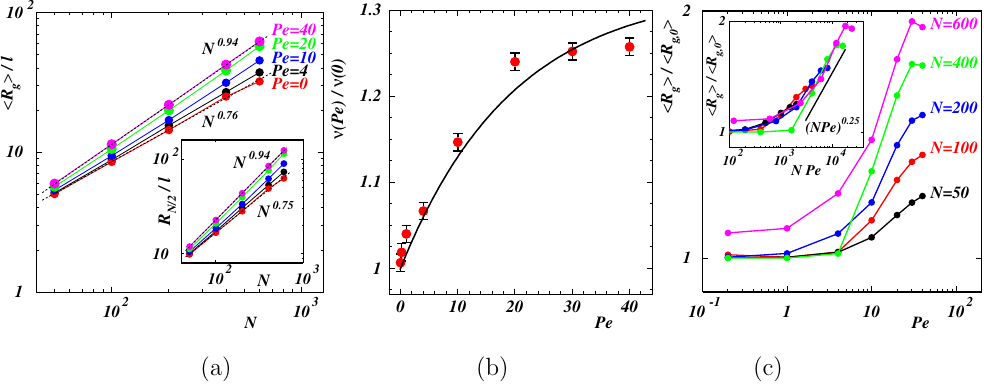}
\caption{Global ring active swelling. a)~Active ring size scaling measured by the mean radius of gyration 
$\lla R_g \rra$ as a function of $N$ for varying activity. Inset: Internal distance $R_{N/2}$ as function of $N$. Symbols are simulation measurements and solid lines correspond to the power law in Eq.~(\ref{eq:radius3}) with $\nu(Pe)$ as fitting parameter.
b)~Activity-dependent Flory exponent $\nu(Pe)$ normalized by the equilibrium value $\nu(0)$. Symbols are the fits to the data as obtained in panel (a) and the solid line corresponds to the growth behavior in Eq.~(\ref{eq:exp}) with $a=0.051$. 
c)~Active ring swelling measured by the mean radius of gyration $\lla R_g \rra/
\lla R_{g,0} \rra$ for polymers of various sizes as a function of $Pe$ in the main plot and of $N Pe$ in the inset. Symbols are simulation measurements and the lines guide connecting the symbols. In the inset the full line corresponds to the trend at intermediate activities $\lla R_g \rra/\lla R_{g,0} \rra \sim (N \mathrm{Pe})^{0.25}$.
\label{fig:scaling}
}
\end{figure*}
Given the variability of the dynamic configurations, we evaluate the probability  distribution functions $P(R_g/\lla R_{g,0}\rra)$ for various values of Pe and $N$ as shown in Fig.~\ref{fig:conf}~d, e. 
For a fixed ring size, Fig.~\ref{fig:conf}~d shows how the relative swelling of the chain with respect to the passive case, clearly increases with activity.
With the accuracy provided by the data, the width of the distributions does not show a significant variation with size or applied activity, which indicates that fluctuations do not change noticeably (cf. Fig.~\ref{fig:conf}~e).

Linear polymers with excluded-volume interactions are well-known to have a power law dependence of their radius of gyration with the polymer contour length, given by the so-called Flory exponent. In two dimensions this exponent has been largely proven to be precisely $\nu_0=3/4$~\cite{norm:92}.  
Simulations in equilibrium and with various activities are performed for rings of different contour lengths and the averaged values of the radius of gyration are shown in Fig.~\ref{fig:scaling}~a. 
Results show very convincingly that the power law dependence applies for rings in equilibrium, and more interestingly also for the cases with an applied tangential activity, being
\begin{equation}
\lla R_g \rra (\mathrm{Pe})\sim l N^{\nu(\mathrm{Pe})}, 
\label{eq:radius3}
\end{equation}
where a generalized activity-dependent exponent $\nu(\mathrm{Pe})$ is considered. 
The passive case is in excellent agreement with the theoretical predictions with  $\nu(0) = 0.76 \pm 0.01$.
 Increasing activity, the swelling scaling exponent $\nu(\mathrm{Pe})$ shows to grow with activity, and in the limit of infinitely large activity, the scaling exponent of rigid circles, $\nu(\infty) =1$, is expected.
 The ratio $\nu(\mathrm{Pe})/\nu(0)$ can therefore be phenomenologically interpolated between the limiting values to increase with the P\'eclet number, by
 
 \begin{equation}
 \frac{\nu(\mathrm{Pe})}{\nu(0)} \simeq 1+ \frac{1-\nu(0)}{\nu(0)} \Big [1-\exp{(-a \mathrm{Pe})} \Big ]. 
 \label{eq:exp} 
 \end{equation}
The numerical data in Fig.~\ref{fig:scaling}~b show to be in reasonable agreement with the fitting value $a=0.051 \pm 0.004$.
The exponent shows to saturate to a constant value at relatively low activities, which  Eq.~(\ref{eq:exp}) captures, with some systematic deviation. 
Similar activity-dependent exponent has also been described in three-dimensional polar rings and polymers~\cite{bianco:18,loca:21,jais:24,jais:26,ayou:26}. However, the dependence with $Pe$ is radically different in those cases since collapsed and non-monotonous behaviors are also observed, which for rings in two dimensions do not occur. 
\begin{figure*}[ht]
\includegraphics*[width=.99\textwidth,angle=0]{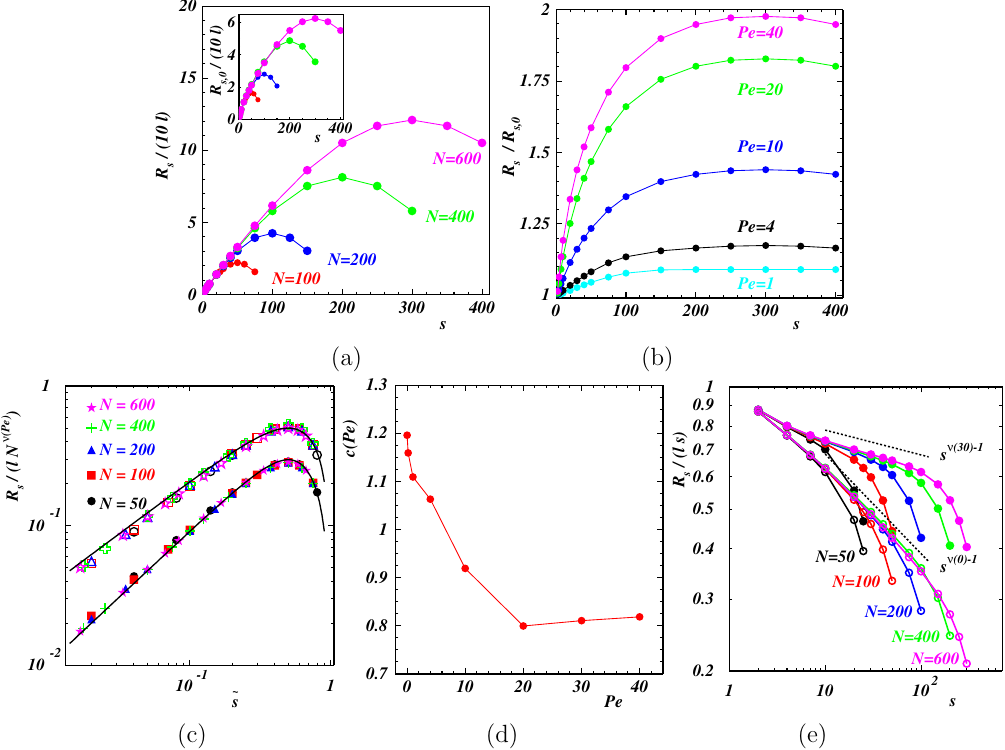}
\caption{Local ring active swelling. a) Average values of the internal distances $R_s$ as functions of the separation $s$, for various values of $N$ with Pe$=30$ (main plot) and in equilibrium (inset).
b)~Relative local swelling, measured by $R_s/R_{s,0}$, with $N=600$ for various Pe values.
c)~Rescaled internal distances $R_{s}$ as functions of the relative separation $\tilde{s}$ in equilibrium (open symbols) and $\mathrm{Pe}=30$ (filled symbols) for various polymer lengths. Lines correspond to a fit with Eq.~(\ref{eq:distance}).
d)~Fitting values $c(\mathrm{Pe})$ in Eq.~(\ref{eq:distance}) as functions of the P\'eclet number.
e)~Internal averaged distances $R_{s}$, rescaled by $l s$, as functions of the separation $s$, for equilibrium (open symbols) and large activity case ($\mathrm{Pe}=30$) (filled symbols). Dashed lines are a guide to the eye for the labeled scaling dependencies. 
\label{fig:distance2pe_noscaling}
}
\end{figure*}

Further insight on the ring swelling behavior is gained by displaying the normalized gyration radii as functions of the applied activity, as shown in Fig.~\ref{fig:scaling}~c.
This representation shows that rings of all sizes significantly swell only in an intermediate range of activities. This can be understood because small values of the active force do not overcome the effect of thermal fluctuations, and for large enough activities, circle-like conformations cannot be further swelled. 
While Fig.~\ref{fig:scaling}~c shows that the range of active forces where the swelling is significant, increases with increasing ring size, the inset shows that all dependencies reasonably collapse into a universal curve when displayed as a function of the so-called global P{\'e}clet number~\cite{teje:24}, this is $N$Pe, which relates to the total force applied in the ring. 
For intermediate activities, the swelling growth shows to be consistent with $\lla R_g \rra / \lla R_{g,0} \rra \sim (N \mathrm{Pe})^{0.25}$, as shown in the inset of Fig.~\ref{fig:scaling}~c.
\mr{On the other hand, it is interesting to note that in the presence of inertia~\cite{faze:23,teje:24}, active polar linear polymers have shown an increase of the radius of gyration at large enough values of the activity, but preceded by a decrease at smaller activities, where inertia is not expected to be relevant. In our simulations, inertia is not present at any value of the activity, and the behavior in Fig.~\ref{fig:scaling}c is fundamentally different since it shows a monotonous increase of the radius of gyration starting already at small values of the activity.}

To characterize the local deformation of rings due to the presence of a tangential active force, we compute the root-mean-squared internal distance 
$R_{s}= \sqrt{\lla (\mathbf{r}_{s+s_0} - \mathbf{r}_{s_0} )^2 \rra}$, which measures the separation length between two beads separated by $s$ bonds.
Due to the closed topology, the property $R_{s}= R_{N-s}$ holds by construction.  
For the half ring case, $s=N/2$, in three dimensions~\cite{baum:82}, it was predicted that $\lla R_g \rra \sim R_{N/2}$. The inset of Fig.~\ref{fig:scaling} a~ clearly confirms that this is also the case for rings in the two-dimensional case. The dependence of internal distances $R_{s}$ with the bead separation $s$ is shown in F$R_{s}= R_{N-s}$a for polymers of various lengths. Due to the closed topology, both passive and active cases show that $R_{s}$ increases with $s$ up to a maximum at $s=N/2$, and also that the increase is larger for larger $N$.
However, the increase is clearly much bigger for the active case than for the passive case, as can also be seen in Fig.~\ref{fig:distance2pe_noscaling}~b, where the active swelling is displayed relative to the passive case.  
Note that the symmetry around $s=N/2$ is due to the closed topology and applies to $R_{s}$, $R_{N-s}$ and $R_{s,0}$,  such that it is observed in Figs.~\ref{fig:distance2pe_noscaling}~a-c.

In the case of passive rings, the functional form of $R_{s}$ is characterized  as a deviation from the Gaussian behavior due to the presence of excluded-volume effects~\cite{bloo:66}, which can be expressed in terms of the Flory exponent, similar to the case of linear polymers. This dependence  has been also confirmed by experiments on circular flexible DNA in two dimensions~\cite{witz:08}. A universal curve was found for the distances $R_{s}$ when normalized by $N^{\nu_0}$ as function of the ratio $\tilde{s}=s/N$ for different polymer lengths. 
\begin{figure*}[ht]
\begin{center}
\includegraphics*[width=.99\textwidth,angle=0]{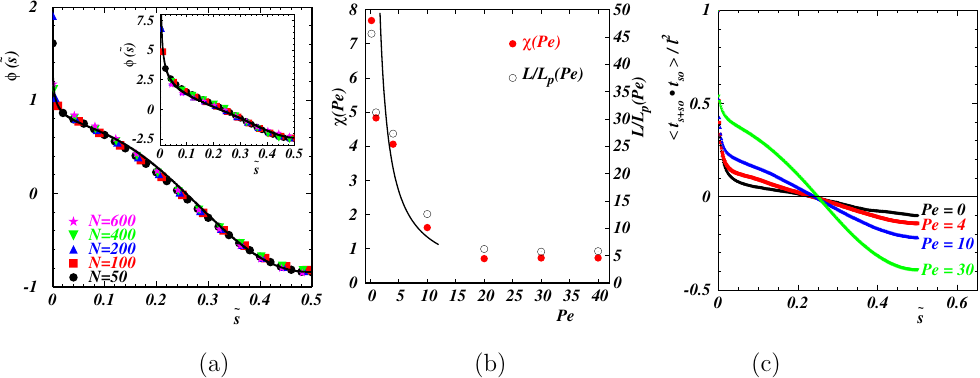}
\caption{a)~Rescaled bond correlation function $\phi(\tilde{s})$ in Eq.~(\ref{eq:tangcorr}) for various polymer lengths, symbols are simulation results and solid lines correspond to Eq.~(\ref{eq:scaling}). Main plot is for the case with Pe$=30$ and the inset is for the equilibrium case. 
b)~Fitting values $\chi(\mathrm{Pe})$ (filled symbols) compared with independent estimations provided by the persistence length $L_p(\mathrm{Pe})$ (empty symbols), both for rings with $L=600 l$. \textcolor{black}{The full line indicates the dependence $L/L_p(Pe) \sim 1/Pe$ as obtained
from Eq.~(\ref{scal}).}
c)~Comparison of the normalized bond correlation functions for rings with $L=600 l$ and varying activity, data shown are from simulation results.
\label{fig:tangcorr}
}
\end{center}
\end{figure*}
\textcolor{black}{In the present system we consider the square root of the Bloomfield--Zimm-type
interpolation for the mean-squared internal distance $R_s^2$ of a passive ring \cite{bloo:66}
replacing the equilibrium exponent $\nu_0$ by the previously calculated activity-dependent exponent, $\nu(Pe)$ and introducing a single fitting
parameter $c(Pe)$,} which is also activity dependent
\begin{equation}
\frac{R_{s}}{l N^{ \nu(\mathrm{Pe})}} = c(\mathrm{Pe}) \frac{\tilde{s}^{ \nu(\mathrm{Pe})} (1-\tilde{s})^{ \nu(\mathrm{Pe})}}{\sqrt{\tilde{s}^{2 \nu(\mathrm{Pe})} + (1-\tilde{s})^{2 \nu(\mathrm{Pe})}}} .
\label{eq:distance}
\end{equation}
\textcolor{black}{The substitution $\nu_0 \rightarrow \nu(Pe)$ was verified to hold
over the whole simulated range $[0,40]$ of the $Pe$ number, observing data collapse for each considered value of $Pe$, as shown in Fig.~\ref{fig:distance2pe_noscaling}~c in two cases. By changing the value
of activity, the fitting parameter $c(Pe)$ was calculated as shown} in Fig.~\ref{fig:distance2pe_noscaling}~d.
The fitting constant decreases from the value at equilibrium $c(\mathrm{0})=1.20\pm 0.01$ to a saturation value $c(\infty) \simeq 0.81 \pm 0.01$ at moderate values of the P\'eclet number. 
The equilibrium value $c(0)$ agrees nicely with the value $1.25$ found in previous numerical simulations of self-excluding semiflexible passive rings in two dimensions~\cite{saka:10}.
Finally, we emphasize that at short separation scales, $\tilde{s} \lesssim 0.1$, long rings behave as linear flexible polymers both in the passive and active case, holding the behavior $R_{s}  \sim l s^{\nu(\mathrm{Pe})}$, as visible in Fig.~\ref{fig:distance2pe_noscaling}~e.

Another important function for understanding the ring conformations  is provided by the bond correlation function $\lla {\bm t}_{s+s_0} \cdot {\bm t}_{s_0} \rra / l^2$, which provides information on the local polymer curvature. 
In the ideal case of a linear Gaussian chain, this correlation function is positively defined and exponentially decays, defining a characteristic polymer persistence length $L_p$. 
Negative correlation values are related to looped configurations, and for rings the function is symmetric with respect to $\tilde{s}=1/2$ by construction.
More complex effects, such as self-avoidance, confinement, specific topologies, or a high polymer concentration, are reflected in different shapes of the bond correlation function, which expresses these constraints in a compact form~\cite{witz:08,witz:11}. 
In equilibrium, it was shown that for ring polymers of different sizes with excluded-volume interactions, the bond correlation function collapses when rescaled considering the Flory exponent~\cite{baum:82}. 
We generalize this result here in the presence of activity by considering  $\nu(\mathrm{Pe})$ as 
\begin{equation}
 \phi(\tilde{s}) \equiv \frac{\lla {\bm t}_{s+s_0} \cdot {\bm t}_{s_0} \rra}{l^2 N^{2\nu(\mathrm{Pe})-2}}
\label{eq:tangcorr}
\end{equation}
The simulation results in Fig.~\ref{fig:tangcorr} show a very clear overlap of the data for five different polymer lengths for the passive case and in one active case. 
For semiflexible passive polymer rings, the scale function  $\phi(\tilde{s})$ has been described as the sum of two contributions~\cite{saka:10}
\begin{equation}
\phi(\tilde{s})=(1-2\tilde{s})\phi_a(\tilde{s})+2\tilde{s}\phi_b(\tilde{s})
\label{eq:scaling}
\end{equation}
with the functions $\phi_a(\tilde{s})$ and $\phi_b(\tilde{s})$ describing the short and long scale behaviors, respectively. 
These functions are here generalized in order to account for the effect of activity, 
\begin{equation}
\phi_a(\tilde{s})= \frac{1}{2}c^2(\mathrm{Pe})\frac{\partial^2}{\partial \tilde{s}^2}
\left [ \frac{\tilde{s}^{2 \nu(\mathrm{Pe})}(1-\tilde{s})^{2 \nu(\mathrm{Pe})}}{\tilde{s}^{2 \nu(\mathrm{Pe})}+(1-\tilde{s})^{2 \nu(\mathrm{Pe})}} \right ]
\label{eq:scaling_S}
\end{equation}
and
\begin{equation}
\phi_b(\tilde{s})=\sqrt{\chi(\mathrm{Pe})}
\left \{ 1+ \frac{\chi(\mathrm{Pe})}{2 \pi^2}\left[g(2 \pi \tilde{s})-g(0)\right] \right \}
\cos(2 \pi \tilde{s})
\label{eq:scaling_L}
\end{equation}
with $g(\tilde{s})=(\pi-\tilde{s})^2/4-\pi^2/12-\cos(\tilde{s})$~\cite{saka:10}.
The value of $\nu(\mathrm{Pe})$ is the same as that calculated for the radius of gyration $R_g$ using Eq.~(\ref{eq:radius3}), and $c(\mathrm{Pe})$ is the same as the one calculated for the internal distances $R_s$  using Eq.~(\ref{eq:distance}). Here, we use $\chi(\mathrm{Pe})$ as the only fitting parameter
\textcolor{black}{as done in the passive case in order to compare to the experimental data 
\cite{saka:10}.}
The fits shown with continuous lines in Fig.~\ref{fig:tangcorr}~a,b excellently capture the behavior of both passive and active flexible rings. 

For semiflexible passive rings the fitting parameter $\chi$ is related to the persistence length as $\chi=L/L_p$. To test this relation, we separately estimate the persistence length for the rings at increasing activity values by fitting the bond correlation function to an exponential decay  $\lla {\bm t}_{s+s_0} \cdot {\bm t}_{s_0} \rra/l^2 \propto \exp(-s/L_p)$ using nearby bonds within the range $\tilde{s}\in (0.002,0.05)$~\cite{rech:09}. The comparison of the obtained $L_p$ values with $\chi(Pe)$ is shown in Fig.~\ref{fig:tangcorr} b. The two functions show to be in a good qualitative agreement, with $\chi(\mathrm{Pe}) \simeq 0.1 L/L_p(\mathrm{Pe})$, over the whole considered range of $Pe$ values.
These persistence fitting parameters $\chi(\mathrm{Pe})$ show to decrease from the value at equilibrium to a saturation constant value at moderately high Pe values, which reflects that the flexible active rings acquire an effective rigidity which increases with activity. 
%\textcolor{red}{This can be understood by the following argument. The c
To support this statement, we consider the adopted active force on each bead $i$
\begin{equation}
|\mathbf{F}_i^a| = f^a \cos\!\left(\frac{\theta_i}{2}\right) \simeq f^a\left(1 - \frac{\theta_i^2}{8}\right),
\end{equation}
with the approximation for small values of the angle $\theta_i$ between consecutive bond vectors.
It is therefore possible to associate an effective local potential 
%can be therefore related with this angular dependence as, 
%%\begin{equation}
$U_{\rm eff}(\theta_i) \approx \text{const} + f^a l \,\theta_i^2 /8 $ 
%%\end{equation}
where $f^a l$ sets the relevant energy scale.
Comparing this with the standard discrete worm-like-chain bending energy,
one can identify the effective bending modulus as, 
%\begin{equation}
$\kappa_{\rm eff} \sim f^a l/4$.
%\end{equation}
By considering  that in two dimensions $L_p =2\kappa/k_{\mathrm{B}}T$, the relation with Pe is obtained
\begin{equation}
\frac{L_p(Pe)}{l} \sim \frac{f^a l}{2 k_BT} = \frac{\mathrm{Pe}}{2}.
\label{scal}
\end{equation}
This equation correctly predicts that, at small activity, the persistence length grows linearly with $\mathrm{Pe}$  (see Fig.~\ref{fig:tangcorr}~b), while the prefactor should be taken as an estimation.
The saturation of $L_p(Pe)$ observed at large $\mathrm{Pe}$ is however outside the estimate since it originates from the finite ring length $L$ and the topological closure constraint on the total curvature. 
%% Having said that, we have to remark a key distinction: active force and effective bending rigidity are not the same physical parameter. The former one is a non-equilibrium driving force while the latter one is a coarse-grained stiffness so that the above argument should be taken with some caution.

This increase of rigidity appears to be similar to other phenomena with increasing rigidity. 
The simulation results of the normalized bond correlation function 
$\lla {\bm t}_{s+s_0} \cdot {\bm t}_{s_0} \rra / l^2$
for various activities are displayed in Fig.~\ref{fig:tangcorr}~c in the case of the longest ring. Here, increasing activity shows how the bond correlation function decays more slowly at short distances followed by a stronger anti-correlation, as observed  in systems with increasing bending rigidity, or due to pressure generated by the excluded volume of enclosed chains~\cite{witz:11}. This is the result of having the larger rigidity coupled to the increase of the overall radius of gyration, namely to observe more circular-like structures.

\section{Summary and conclusions} \label{sec:conclusions}

Polar ring polymers adsorbed on a surface show to swell with increasing applied activity, in contrast with three-dimensional polar active ring polymers which show collapsed structures for high enough values of the activity. This difference is known to be due to the dimensionality of the system~\cite{lamu:24} similar to that observed also in self-propelled polymers which display spontaneous spiral formation only in the two-dimensional case~\cite{isele:15}. The well-known scaling laws of 
cyclic filaments in equilibrium can be generalized for the conformational properties of two-dimensional polar active ring polymers by considering activity-dependent coefficients. The radius of gyration follows a power-law increase with the ring length, and the growth exponent increases with activity. In the limit of low  activity, the exponent tends to the standard Flory exponent of passive systems, and for very large activities it saturates to 1, characteristic of a perfect circle configuration. 
The local extension of the polymer segment follows the same behavior as that in equilibrium  when considering the activity-dependent Flory exponent together with an activity-dependent prefactor. 
Furthermore, the bond correlation function can be described by the sum of short- and long-scale behaviors generalizing the equilibrium case, with an additional single fitting parameter that relates to an effective persistence length. Increasing activity has the same effect in the bond correlation function as augmenting the bending rigidity or the pressure generated by the excluded volume of enclosed chains.
% Having said that, we have to remark a key distinction: active force and effective bending rigidity are not the same physical parameter. The former one is a non-equilibrium driving force while the latter one is a coarse-grained stiffness so that the above argument should be taken with some caution.
Although related, active force and effective bending rigidity have a different physical origin. While the effective bending due to activity is a non-equilibrium driving force which might dynamically vary, the bending due to rigidity is a coarse-grained effect due to the specific polymer composition which remains over time but might vary locally. 
These results establish universal swelling as a robust feature of tangentially driven ring polymers in two dimensions and highlight the combined role of dimensionality, excluded-volume interactions, and topology in determining the scaling behavior of active polymer systems.

\acknowledgments

The work of AL was performed under the auspices of GNFM-INdAM.

%\appendix*

\bibliography{polymer_bib}

\end{document}